\documentclass[conference]{IEEEtran}

\usepackage{amsmath}
\usepackage{booktabs}
\usepackage{graphicx}
\usepackage[table]{xcolor}
\usepackage{url}
\definecolor{qtwoshade}{RGB}{226,244,224}
\definecolor{qthreeshade}{RGB}{224,238,250}
\newcommand{\qtwowin}[1]{\cellcolor{qtwoshade}#1}
\newcommand{\qthreewin}[1]{\cellcolor{qthreeshade}#1}
\newcommand{\qtwopart}[1]{{\setlength{\fboxsep}{0.4pt}\colorbox{qtwoshade}{#1}}}
\newcommand{\qthreepart}[1]{{\setlength{\fboxsep}{0.4pt}\colorbox{qthreeshade}{#1}}}

\title{Cached LLM Probability Retrieval for Speech Recognition}

\author{
\IEEEauthorblockN{Sheng Li}
\IEEEauthorblockA{\textit{Institute of Science Tokyo}\\\textit{Kyoto University}\\
Yokohama, Kyoto, Japan}
\and
\IEEEauthorblockN{Takahiro Shinozaki}
\IEEEauthorblockA{\textit{Institute of Science Tokyo }\\\textit{Yokohama, Japan}\\
}
\and
\IEEEauthorblockN{Tatsuya Kawahara}
\IEEEauthorblockA{\textit{Kyoto University}\\\textit{Kyoto, Japan}\\
}
}

\begin{document}
\maketitle

\begin{abstract}
Large language models (LLMs) enhance automatic speech recognition (ASR) by providing linguistic priors; however, their direct rescoring is costly because it requires evaluating every N-best hypothesis. This paper introduces "cached LLM probability retrieval," which involves querying a local teacher LLM offline to obtain next-token probabilities for ASR-relevant context-target pairs. These probabilities are then utilized during recognition via cache lookups, backoff strategies, and optional scoring for significant misses. The method is training-free and can integrate with existing recognizers without requiring modifications to acoustic models. Evaluations across various ASR models reveal that cached retrieval outperforms 1-pass ASR in 28 of 39 settings and achieves lower non-oracle errors. Context length analysis indicates that benefits peak at a context length of 8, suggesting that cached probability retrieval is an effective and lightweight ASR adaptation method, in contrast to the heavy training required for Generative Error Correction (GER) or knowledge distillation (KD).
\end{abstract}


\section{Introduction}

\IEEEPARstart{M}{odern} ASR systems are increasingly used as fixed foundation recognizers: Whisper models trained with large-scale weak supervision~\cite{radford2022whisper}, self-supervised speech encoders such as wav2vec~2.0~\cite{baevski2020wav2vec2} and HuBERT~\cite{hsu2021hubert}, and toolkit systems such as SpeechBrain~\cite{ravanelli2021speechbrain}. In many deployment settings, the user of the recognizer cannot retrain the acoustic model or rebuild the decoder. A practical adaptation method should therefore operate on hypotheses that are already available, such as an N-best list, beam, or lattice, and should add a useful language prior with minimal additional training.

LLMs are attractive for this role because they assign high probabilities to fluent, semantically plausible text. However, directly placing an LLM in the ASR rescoring loop is costly: each hypothesis requires evaluating its probability autoregressively, which adds latency, memory pressure, and deployment complexity. Finetuned generative error correction (GER) and knowledge distillation can reduce this cost at runtime, but they require paired supervision, parameter training, and an additional adaptation pipeline. These costs motivate a complementary question: can an LLM be converted into a reusable local scoring resource before recognition?

This paper answers that question with cached LLM probability retrieval. Offline, a local causal LLM scores context-target token pairs drawn from ASR-relevant non-test text. Online, an existing recognizer produces an N-best list, and each hypothesis receives an additive cached LLM score by lookup. Missing entries are handled by shorter-context backoff; a selective policy can call the teacher LLM only for ambiguous high-impact misses. The fully local policy makes no online LLM calls after cache construction, while the selective policy uses online LLM computation only where it can plausibly change the ranking.
The usefulness of this design is its position between classical LM rescoring and trained LLM correction. Like an $n$-gram or cache LM, it is local, transparent, and easy to attach to an existing recognizer. Unlike empirical count caches, it stores teacher LLM probabilities, so the retrieved score encodes a stronger language prior than raw frequency alone. Unlike LoRA-GER or KD, it requires no parameter update for the rescoring model. These properties make it suitable as a lightweight adaptation layer, a diagnostic for whether LLM probabilities can rank a given N-best space, and a component that can be combined with stronger trained correction models.

Our contributions are threefold. First, we formulate cached LLM probability retrieval for ASR N-best rescoring, including local backoff and selective direct LLM scoring for important cache misses. Second, we evaluate the method across six recognizer families and multiple datasets/languages, and compare it with Qwen2.5/Qwen3 LoRA-GER baselines on the same N-best lists. Third, we identify the operating regime in which the method is most useful: cached retrieval improves 28 of 39 full-data experimental settings and all Whisper-small settings, but its benefit depends on N-best diversity and cache coverage. A full context sweep further shows that short cache contexts are sufficient in this exact-lookup setting, making the method practical to cache and deploy.

\begin{figure*}[t]
    \centering
    \includegraphics[width=1\textwidth]{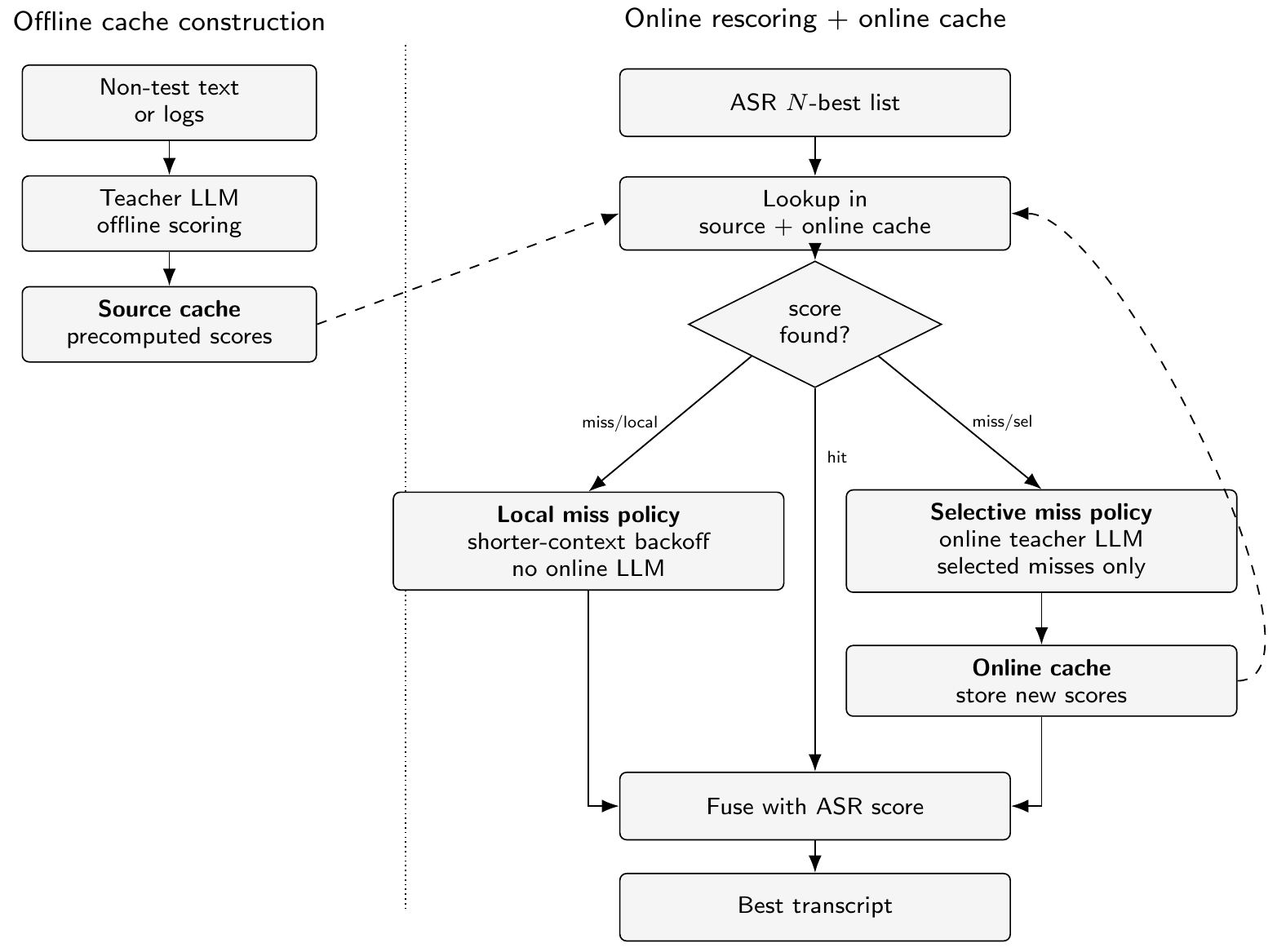}
    \caption{Overview of proposed method: offline caching and online rescoring (with two backoff strategies)}
    \label{fig:framework}
\end{figure*}

\section{Related Work}

\subsection{ASR Language Model Rescoring}

Classical $n$-gram language models remain widely used in ASR because they are compact, fast, and easy to combine with beam search~\cite{chen1999smoothing,heafield2011kenlm}. Neural language models improve fluency modeling but increase inference cost~\cite{mikolov2010rnnlm}. N-best rescoring is a compromise: the recognizer proposes a set of candidates, and an external model reranks them. The oracle error rate of the N-best list is therefore important, since no rescoring method can recover hypotheses that are not present.

Retrieval and cache language models store information from previous or external text and retrieve it during scoring~\cite{khandelwal2020knnlm}. Our method differs in the stored value. We do not store counts or nearest-neighbor embeddings. Instead, we store a teacher LLM probability $\log p_\theta(y_t \mid c_t)$ for an ASR hypothesis context $c_t$ and target token $y_t$.

\subsection{LLM Knowledge Transfer for ASR}

KD transfers knowledge from a larger teacher to a smaller student~\cite{hinton2015distilling}. In ASR, BERT or LLM information has been distilled into sequence-to-sequence and CTC systems~\cite{futami2020distilling,futami2022ctc}, and ASR-specific distillation has also been used to compress strong recognizers~\cite{gandhi2023distilwhisper}. More broadly, LLM representations can be transferred into ASR systems~\cite{udagawa2023multiple}, and multiple foundation speech models can act as teachers for ASR students~\cite{yang2023foundationkd}. These methods train parameters and can generalize beyond memorized contexts.

Cached LLM probability retrieval is adjacent to KD but not identical. The teacher LLM is used offline, but no student is trained. This makes the runtime system deterministic and inspectable: a score exists if a key exists, otherwise the system backs off or selectively calls the teacher. The tradeoff is that exact lookup has weaker generalization than a trained student model.

\section{Method}
The proposed method is shown in Figure \ref{fig:framework}.
\subsection{Cached Probability Scoring}

Let $\mathcal{H}(x)=\{y^{(1)},\ldots,y^{(N)}\}$ be an N-best list for utterance $x$. Each hypothesis is tokenized by the LLM tokenizer as $y=(y_1,\ldots,y_T)$. For maximum context length $K$, the cached LLM score is
\begin{equation}
S_K(y)=\sum_{t=1}^{T} \log p_\theta(y_t \mid y_{\max(1,t-K):t-1}),
\end{equation}
where $\theta$ is a local causal LLM. We store both summed and averaged token log probabilities. In a full-prefix diagnostic score, $K$ is allowed to cover all previous tokens in the utterance.

Each cache key consists of the context token sequence and the target token id. In a research diagnostic setting, keys can be extracted from the evaluation N-best list to ask whether the LLM probability contains useful ranking information for the observed hypotheses. In a deployable setting, keys must be constructed from train, development, domain, or logged text before test-time recognition.

\subsection{Backoff and Selective Substitution}

For a missing key at context length $K$, the system backs off to shorter context lengths:
\begin{equation}
K \rightarrow K/2 \rightarrow \cdots \rightarrow 1.
\end{equation}
If all contexts miss, a neutral fallback score is used. This policy is fully local and uses no runtime LLM call.

We also evaluate a selective substitution policy. For an utterance, if the current score margin between the top two hypotheses is smaller than a threshold $\tau$, direct LLM scoring is allowed for missing context-target pairs in the top-$M$ hypotheses. Otherwise, the system uses ordinary backoff. The idea is to reserve expensive LLM calls for ambiguous cases where the ranking can actually change.

\subsection{Retrieval-Adapted Local Scores}

For earlier diagnostic comparisons, we also consider a local retrieval $n$-gram feature. The first ASR hypothesis is used as a query over non-test text from the same corpus or language. The top retrieved snippets define an utterance-specific smoothed 5-gram model. This feature captures local domain style and vocabulary. It is not an LLM and is deliberately lightweight.

\subsection{LoRA-Finetuned GER Baseline}

To compare against a stronger parameter-trained LLM baseline, we fine-tune Qwen2.5-0.5B and Qwen3.0-8B with LoRA~\cite{hu2022lora} for generative error correction. The supervised input is a prompt containing an ASR transcript, and the target is the reference transcript. Training uses only non-test LibriSpeech dev-other material: the union of the six ASR models' derived dev N-best lists, plus dev-other reference-copy examples that regularize the model against unnecessary rewriting. The Qwen2.5 adapter has rank 8 and is trained for one epoch. We also train a local Qwen3-8B LoRA adapter with 4-bit loading, rank 4, and one epoch. Both adapters are evaluated on all available full-data N-best experimental settings for the six ASR models.

Free-form GER can produce strings outside the recognizer's N-best list and can introduce hallucinated corrections. Since it is still one of the strongest N-best rescoring methods, we use GER as strong baseline. 

\subsection{Fusion}

All rescoring features are standardized within a corpus and interpolated:
\begin{equation}
\hat{y}=\arg\max_{y\in\mathcal{H}(x)}
\lambda_a z_a(y)+\lambda_q z_q(y)+\lambda_r z_r(y),
\end{equation}
where $z_a$, $z_q$, and $z_r$ are the ASR score, cached LLM score, and retrieval $n$-gram score when the retrieval feature is enabled. In the leakage-free tables, weights are selected on a development split.

\section{Experimental Setup}

\subsection{Datasets and Metrics}

We evaluate LibriSpeech test-other~\cite{panayotov2015librispeech}, a 5dB noise-mixed version of LibriSpeech test-other, AMI IHM meeting speech~\cite{carletta2005ami}, and FLEURS~\cite{conneau2022fleurs}. For FLEURS, use WER for English (en), German (de), Spanish (es), and French (fr). Chinese Mandarin (cmn) and Japanese (ja) use CER as the metric.

\subsection{ASR Model Recognizers and Teacher LLMs}

The Whisper family includes Whisper-base, Whisper-small, and Whisper-large-v3. The self-supervised CTC family uses wav2vec~2.0 base and HuBERT-large English recognizers, and is evaluated on LibriSpeech test-other, noisy LibriSpeech test-other (5 dB noise mixing), AMI IHM, and FLEURS English. The SpeechBrain family uses a CRDNN/RNNLM system with its beam-search N-best output. For Whisper family models, evaluations include multilingual speech recognition, while other models' evaluations are English-only.

The main teacher comparison uses local Qwen2.5-0.5B~\cite{yang2024qwen25} and local Qwen3-8B~\cite{yang2025qwen3}. 

\subsection{Other Experimental Settings}

We report a leakage-free source-cache protocol. Cache entries are built only from non-test text, such as development, training, or language-matched FLEURS text. At test time, the local policy uses exact lookup and only shorter-context backoff. When a separate development N-best list is available, interpolation weights are tuned on it; otherwise, we use a non-overlapping dev-test split.

The LoRA-GER baseline for each LLM teacher uses an adapter trained on ASR-reference supervision and N-best ASR hypotheses. We report results on the LibriSpeech test-other, noisy LibriSpeech test-other (5 dB noise mixing), AMI, and FLEURS. 

For context sensitivity, we repeat the source-cache protocol for $K\in\{8,16,32,64,128,256,512,1024\}$ across both teacher LLMs, all ASR models, and all available full-data datasets. Table~\ref{tab:qwen_teacher_full} reports both the fully local shorter-context backoff policy and a selective direct-scoring policy for important misses in the $K=32$ policy-comparison runs; Table~\ref{tab:context_matrix} isolates the fully local backoff-only context sweep so that $K$ is the only changing variable.

\section{Results}

\subsection{Teacher LLM, ASR models, and Cache Coverage Scaling}

Table~\ref{tab:qwen_teacher_full} is the main usefulness test for the proposed cached retrieval score. It compares Qwen2.5-0.5B and Qwen3-8B on full N-best lists from Whisper-base, Whisper-small, Whisper-large-v3, SpeechBrain CRDNN/RNNLM, wav2vec~2.0 base, and HuBERT-large. All rows use maximum cache context $K=32$ with shorter-context backoff over $\{16,8,4,2,1\}$. For each teacher, ``local'' is source-cache/backoff-only scoring, while ``sel'' additionally allows selective direct LLM scoring for important cache misses. The H/D/M columns report source-cache hit, selective-direct, and miss case percentages.

The table shows that cached LLM probabilities are useful as an additive rescoring feature in a large fraction of conditions. Taking the best local/sel cache score across the two teachers, cached retrieval improves first-pass ASR in 28 of 39 ASR model-dataset settings and is tied for the lowest non-oracle error in 25 of 39 settings. The effect is strongest for Whisper-small: all nine Whisper-small settings improve, with an average best-cache reduction of 8.13\% absolute WER/CER. The largest gains are 13.01\% WER on AMI IHM, 9.22\% CER on Mandarin FLEURS, 8.86\% WER on noisy LibriSpeech, and 7\% to 9\% on several European/Japanese FLEURS settings. Cached retrieval is also consistently positive on the self-supervised CTC systems: all wav2vec~2.0 and HuBERT rows improve over first-pass decoding (1-pass).

We also observed Whisper-large-v3 already ranks many N-best lists well, so little recoverable text-only room remains. SpeechBrain CRDNN/RNNLM is less responsive to the cached score, especially on LibriSpeech and noisy LibriSpeech. These negative or flat settings are useful evidence rather than a failure of the idea: they show that cached LLM retrieval is a targeted adaptation, most valuable when the recognizer produces plausible alternatives that are linguistically distinguishable and when cache/backoff coverage is adequate. The colored H/D/M entries make this mechanism explicit: local wins correlate with source-cache hits/backoff, while selective wins are associated with direct LLM scoring of important misses.

\begin{table*}[t]
\centering
\caption{Teacher LLM, ASR models, two types of cached-retrieval (local and sel), and LoRA-GER (GER) comparison with maximum cache context $K=32$. }
\label{tab:qwen_teacher_full}
\begin{tabular}{llrrrrrrrrrr@{\hspace{0.8em}}l}
\toprule
ASR model & Dataset & 1-pass & \multicolumn{4}{c}{Qwen2.5-0.5B (Q2) ASR err (\%) } & \multicolumn{4}{c}{Qwen3-8B (Q3) ASR err (\%)} & Oracle & Metric \\
\cmidrule(lr){4-7}\cmidrule(lr){8-11}
 & & err(\%) & local & sel & GER & H/D/M rate(\%) & local & sel & GER & H/D/M rate(\%) & err(\%) & \\
\midrule
whisper-base & Test-other(LS) & 21.01 & 22.03 & 22.05 & 20.27 & 20.3/3.9/75.8 & 21.47 & 21.53 & \textbf{20.15} & 20.0/2.4/77.6 & 10.10 & WER \\
   & LS 5dB mix & \textbf{55.31} & 59.58 & 59.46 & 58.07 & 23.3/2.8/73.9 & 58.78 & 58.78 & 58.01 & 22.2/1.1/76.7 & 38.25 & WER \\
   & AMI IHM & \textbf{42.68} & 46.11 & 46.90 & 64.89 & 1.6/16.8/81.6 & 46.10 & 46.66 & 64.17 & 1.3/16.1/82.6 & 18.89 & WER \\
   & FLEURS en & 19.53 & \qtwowin{\textbf{16.22}} & 16.24 & 17.97 & \qtwopart{20.6}/6.4/73.0 & 16.79 & 16.78 & 17.74 & 20.2/7.9/71.9 & 9.92 & WER \\
   & FLEURS cmn & 50.98 & 41.14 & \qtwowin{\textbf{41.10}} & 58.55 & 2.6/\qtwopart{0.8}/96.6 & 41.33 & 41.48 & 58.45 & 2.4/0.6/97.0 & 30.12 & CER \\
   & FLEURS de & 36.69 & 32.23 & \qtwowin{\textbf{31.97}} & 34.71 & 11.3/\qtwopart{5.3}/83.3 & 32.39 & 32.19 & 34.29 & 11.1/6.0/82.9 & 22.53 & WER \\
   & FLEURS es & 22.48 & 18.33 & 18.29 & 18.04 & 19.7/5.4/74.9 & 18.59 & 18.44 & \textbf{17.68} & 19.0/7.1/73.9 & 10.66 & WER \\
   & FLEURS fr & 46.83 & 40.48 & \qtwowin{\textbf{40.37}} & 49.51 & 18.9/\qtwopart{9.6}/71.5 & 40.84 & 40.61 & 49.22 & 19.3/4.3/76.4 & 28.38 & WER \\
   & FLEURS ja & 47.67 & 38.74 & \qtwowin{\textbf{38.56}} & 52.18 & 14.6/\qtwopart{4.3}/81.1 & 40.56 & 40.50 & 52.07 & 14.4/0.5/85.1 & 27.89 & CER \\
\midrule
whisper-small & LS & 15.32 & 9.72 & \qtwowin{\textbf{9.70}} & 10.24 & 20.9/\qtwopart{2.5}/76.6 & 9.76 & 9.76 & 9.75 & 20.7/2.5/76.8 & 6.15 & WER \\
   & LS 5dB mix & 38.95 & \qtwowin{\textbf{30.09}} & 30.11 & 31.92 & \qtwopart{21.8}/1.2/77.0 & 30.30 & 30.30 & 30.13 & 21.5/1.9/76.6 & 23.14 & WER \\
   & AMI IHM & 37.39 & 24.40 & 25.22 & 26.79 & 1.4/9.8/88.7 & \qthreewin{\textbf{24.38}} & 25.32 & 25.20 & \qthreepart{1.4}/6.3/92.2 & 16.03 & WER \\
   & FLEURS en & 14.29 & 10.81 & 10.94 & 10.37 & 21.7/6.8/71.5 & 10.52 & 10.52 & \textbf{9.68} & 21.7/5.9/72.3 & 6.87 & WER \\
   & FLEURS cmn & 34.01 & \qtwowin{\textbf{24.79}} & 24.88 & 29.25 & \qtwopart{2.5}/1.4/96.1 & 25.12 & 25.20 & 28.19 & 2.2/2.3/95.5 & 16.61 & CER \\
   & FLEURS de & 22.78 & 14.80 & 14.78 & 15.67 & 12.5/3.7/83.8 & 14.71 & \qthreewin{\textbf{14.69}} & 14.79 & 12.5/\qthreepart{3.8}/83.6 & 11.50 & WER \\
   & FLEURS es & 13.79 & 6.53 & 6.53 & 7.58 & 21.2/5.8/73.0 & \qthreewin{\textbf{6.51}} & \qthreewin{\textbf{6.51}} & 6.70 & \qthreepart{20.7}/\qthreepart{8.7}/70.6 & 4.57 & WER \\
   & FLEURS fr & 26.84 & 17.46 & 17.44 & 18.28 & 22.0/4.1/73.8 & 17.46 & 17.39 & \textbf{17.30} & 21.8/5.2/73.1 & 13.38 & WER \\
   & FLEURS ja & 25.12 & 17.34 & 17.35 & 19.14 & 17.3/0.3/82.3 & 17.29 & \qthreewin{\textbf{17.28}} & 17.59 & 17.0/\qthreepart{1.5}/81.5 & 12.69 & CER \\
\midrule
whisper-large-v3 & LS & \textbf{4.34} & 4.66 & 4.67 & 4.63 & 26.6/3.4/70.0 & 4.46 & 4.45 & 4.52 & 24.7/2.8/72.5 & 2.46 & WER \\
   & LS 5dB mix & \textbf{12.72} & 13.36 & 13.36 & 12.86 & 23.2/5.5/71.4 & 13.02 & 13.02 & 12.86 & 21.7/7.1/71.2 & 9.17 & WER \\
   & AMI IHM & 21.32 & \qtwowin{\textbf{21.30}} & \qtwowin{\textbf{21.30}} & 21.98 & \qtwopart{1.3}/\qtwopart{19.2}/79.6 & 21.32 & 21.39 & 21.74 & 1.2/10.5/88.4 & 13.90 & WER \\
   & FLEURS en & \textbf{6.22} & 6.25 & 6.24 & \textbf{6.22} & 22.5/2.4/75.1 & 6.23 & \qthreewin{\textbf{6.22}} & 6.31 & 22.6/\qthreepart{1.2}/76.2 & 4.70 & WER \\
   & FLEURS cmn & 8.28 & \qtwowin{\textbf{8.12}} & 8.18 & 8.28 & \qtwopart{2.5}/3.3/94.2 & 8.29 & 8.28 & 8.22 & 2.3/5.8/91.8 & 6.47 & CER \\
   & FLEURS de & 7.77 & 7.82 & \qtwowin{\textbf{7.69}} & 7.79 & 13.3/\qtwopart{2.5}/84.2 & 7.83 & 7.81 & \textbf{7.69} & 13.2/2.1/84.6 & 5.66 & WER \\
   & FLEURS es & 2.72 & \qtwowin{\textbf{2.71}} & \qtwowin{\textbf{2.71}} & 2.77 & \qtwopart{22.1}/\qtwopart{0.4}/77.5 & 2.72 & 2.72 & 2.72 & 22.1/0.7/77.3 & 1.70 & WER \\
   & FLEURS fr & \textbf{6.38} & 6.45 & 6.45 & \textbf{6.38} & 22.2/4.5/73.3 & \qthreewin{\textbf{6.38}} & 6.39 & 6.52 & \qthreepart{22.7}/2.1/75.2 & 3.96 & WER \\
   & FLEURS ja & \textbf{4.41} & 4.47 & 4.43 & \textbf{4.41} & 17.1/9.3/73.6 & 4.43 & 4.42 & \textbf{4.41} & 18.2/1.9/80.0 & 2.71 & CER \\
\midrule
CRDNN/RNNLM & LS & 9.92 & 11.43 & 11.39 & 9.91 & 31.3/4.3/64.4 & 10.57 & 10.47 & \textbf{9.87} & 30.5/6.0/63.5 & 6.69 & WER \\
   & LS 5dB mix & 32.21 & 33.16 & 33.14 & \textbf{32.19} & 30.7/6.4/62.9 & 32.85 & 32.82 & \textbf{32.19} & 29.8/8.0/62.2 & 27.61 & WER \\
   & AMI IHM & 56.16 & \qtwowin{\textbf{56.02}} & 56.32 & 56.05 & \qtwopart{0.0}/52.9/47.1 & \qthreewin{\textbf{56.02}} & 56.77 & 56.12 & \qthreepart{0.0}/52.2/47.8 & 48.80 & WER \\
   & FLEURS en & 30.92 & 30.92 & 30.92 & 30.90 & 3.8/87.9/8.3 & 30.92 & 30.92 & \textbf{30.71} & 3.8/87.9/8.3 & 27.27 & WER \\
\midrule
wav2vec2-base & LS & 8.45 & 8.39 & 8.39 & 8.39 & 31.3/0.4/68.3 & 8.48 & 8.48 & \textbf{8.27} & 31.1/0.3/68.6 & 6.50 & WER \\
   & LS 5dB mix & 76.90 & 76.66 & \qtwowin{\textbf{76.62}} & 76.89 & 15.1/\qtwopart{14.5}/70.5 & 76.72 & 76.70 & 76.87 & 14.4/17.7/67.8 & 75.61 & WER \\
   & AMI IHM & 41.85 & \qtwowin{\textbf{41.09}} & 41.29 & 41.68 & \qtwopart{0.0}/24.4/75.6 & \qthreewin{\textbf{41.09}} & 41.43 & 41.65 & \qthreepart{0.0}/12.4/87.6 & 38.42 & WER \\
   & FLEURS en & 21.29 & \qtwowin{\textbf{20.57}} & \qtwowin{\textbf{20.57}} & 21.22 & \qtwopart{23.0}/\qtwopart{8.2}/68.8 & 20.71 & 20.69 & 20.98 & 22.1/11.1/66.8 & 19.01 & WER \\
\midrule
hubert-large & LS & 4.16 & 4.13 & 4.14 & 4.14 & 31.6/0.7/67.7 & 4.17 & 4.17 & \textbf{4.04} & 31.6/0.9/67.6 & 2.71 & WER \\
   & LS 5dB mix & 22.96 & \qtwowin{\textbf{22.62}} & 22.63 & 22.86 & \qtwopart{27.8}/3.8/68.4 & 22.77 & 22.77 & 22.74 & 27.3/5.3/67.4 & 21.08 & WER \\
   & AMI IHM & 35.83 & \qtwowin{\textbf{35.19}} & 35.34 & 35.60 & \qtwopart{0.0}/23.4/76.6 & \qthreewin{\textbf{35.19}} & 35.63 & 35.59 & \qthreepart{0.0}/11.9/88.1 & 32.35 & WER \\
   & FLEURS en & 15.26 & \qtwowin{\textbf{14.92}} & \qtwowin{\textbf{14.92}} & 15.20 & \qtwopart{25.0}/\qtwopart{4.4}/70.6 & 15.10 & 15.03 & 15.01 & 23.8/8.4/67.8 & 13.17 & WER \\
\bottomrule
\\
\end{tabular}
\begin{itemize}
  \item Values are primary error rates: WER, except for Mandarin and Japanese FLEURS, where CER is used. 
  \item Results are grouped by LLM teacher: the Q2 block reports Qwen2.5-0.5B local, sel, GER, and H/D/M values, and the Q3 block reports the corresponding Qwen3-8B values. Local is source-cache/backoff-only scoring; sel allows selective direct scoring for important misses. GER is the N-best LoRA-GER baseline. H/D/M gives source-cache hit, selective-LLM-direct, and miss token percentages. 
  \item Bold marks the lowest non-Oracle error in each row; a lower error is better. Light green and light blue mark Q2 and Q3 local/sel scores only when they are tied for the row-best non-oracle error.
  \item Within H/D/M, only H is shaded for a local win, and only D is shaded for a sel win; M is not shaded.
\end{itemize}
\end{table*}

After integrating the LoRA-GER columns, Table~\ref{tab:qwen_teacher_full} also separates two different kinds of LLM use. LoRA-GER is a trained correction prior, while cached retrieval is a lookup-based probability prior. GER is strongest in several rows, especially where Qwen3-8B benefits from supervised adaptation, but it does not dominate the table. Cached local/sel scores are tied for the row-best non-oracle error more often than the GER columns in this N-best comparison, while requiring no parameter training. The practical message is, therefore, not that one mechanism replaces the other, but that cached probabilities provide a useful, low-cost LLM feature that can be used when fine-tuning is unavailable or combined with trained correction when supervision exists.

\subsection{LoRA-GER Strong Baseline}

The LoRA-GER baselines in Table~\ref{tab:qwen_teacher_full} provide a strong reference point because they use parameter training but are evaluated under the same N-best oracle bound as cached retrieval: the GER model must choose among recognizer hypotheses rather than generate an arbitrary new sentence. This comparison strengthens the interpretation of the proposed method. If a trained GER system is available and matched to the ASR errors, it can be effective. A larger LLM-based GER usually outperforms a smaller LLM-based GER. For example, in the Whisper-small setting, Qwen3-8B GER improves over Qwen2.5-0.5B GER by 0.49\% WER on LibriSpeech, 1.79\% WER under 5 dB noise, 1.59\% WER on AMI, and 1.55\% CER on Japanese FLEURS.

At the same time, Table~\ref{tab:qwen_teacher_full} shows that the training-free cached score remains useful under this stronger comparison. Cached retrieval gives the best non-oracle score in many Whisper-base, Whisper-small, wav2vec~2.0, and HuBERT rows and avoids the paired-data requirement and adaptation cost of LoRA-GER. The comparison therefore positions cached LLM retrieval as a lightweight LLM adaptation method; it is not meant to replace GER when supervised fine-tuning is feasible, but it can deliver substantial gains without that fine-tuning step and can serve as an interpretable companion feature to trained correction.

\subsection{Context Window Length}

Table~\ref{tab:context_matrix} reports the full backoff-only context-sensitivity matrix over the same teacher LLM-ASR model-dataset coverage as Table~\ref{tab:qwen_teacher_full}. For each row, we sweep $K\in\{8,16,32,64,128,256,512,1024\}$ while keeping the source cache, N-best list, tuning split, and local backoff rule fixed. The result is an efficiency advantage for the proposed design: no row prefers $K>32$. In 67 of 78 rows of different settings, $K=8$ is already the best or tied-best setting; six rows prefer $K=16$, and five prefer $K=32$. The maximum absolute difference between $K=32$ and any longer context is 0.004\% WER/CER.

\begin{table*}[t]
\centering
\caption{Full context-sensitivity matrix over two teacher LLMs, six ASR models, and all available full-data datasets. Best-K counts show which context length minimizes the primary error metric within a teacher LLM-ASR model group. $\Delta_{32}$ is the mean change from first-pass ASR at $K=32$; negative is better. Long-gap is $\max_{K>32} |E_K-E_{32}|$.}
\label{tab:context_matrix}
\begin{tabular}{llrllrrr}
\toprule
Teacher LLM & ASR models & \#Datasets & Best-$K$ counts & $\Delta_{32}$ & Mean hit & Hit range & Long-gap (err\%) \\
\midrule
Qwen2.5-0.5B & HuBERT-large & 4 & 8:2, 16:2 & -0.301 & 21.51 & 0.00--31.72 & 0.000 \\
Qwen3-8B & HuBERT-large & 4 & 8:3, 32:1 & -0.235 & 21.43 & 0.00--31.72 & 0.000 \\
Qwen2.5-0.5B & CRDNN/RNNLM & 4 & 8:4 & 0.575 & 22.17 & 0.00--32.36 & 0.004 \\
Qwen3-8B & CRDNN/RNNLM & 4 & 8:4 & 0.300 & 21.98 & 0.00--32.05 & 0.000 \\
Qwen2.5-0.5B & wav2vec~2.0 & 4 & 8:1, 16:2, 32:1 & -0.416 & 18.00 & 0.00--31.31 & 0.000 \\
Qwen3-8B & wav2vec~2.0 & 4 & 8:4 & -0.356 & 17.84 & 0.00--31.20 & 0.000 \\
Qwen2.5-0.5B & Whisper-base & 9 & 8:7, 32:2 & -3.683 & 15.32 & 1.18--23.54 & 0.000 \\
Qwen3-8B & Whisper-base & 9 & 8:7, 16:2 & -3.112 & 14.57 & 1.03--21.46 & 0.000 \\
Qwen2.5-0.5B & Whisper-large-v3 & 9 & 8:9 & 0.135 & 17.21 & 1.27--27.01 & 0.000 \\
Qwen3-8B & Whisper-large-v3 & 9 & 8:9 & 0.148 & 17.07 & 1.22--26.38 & 0.000 \\
Qwen2.5-0.5B & Whisper-small & 9 & 8:8, 32:1 & -8.024 & 16.06 & 1.37--22.88 & 0.000 \\
Qwen3-8B & Whisper-small & 9 & 8:9 & -7.959 & 15.98 & 1.29--22.79 & 0.000 \\
\bottomrule
\end{tabular}
\end{table*}

This result is important for deployment. It does not claim that long LLM histories are never useful for ASR; rather, it shows that exact lookup changes the economics of context. Longer contexts create many more cache keys, but most N-best utterances can already be distinguished by short local histories or by the backoff path. Therefore, the useful cached feature can be built with a small context cap, reducing cache construction cost and lookup sparsity. Long-context evidence should be reserved for selective direct scoring or semantic retrieval in rare ambiguous cases, instead of being stored exhaustively for every token.

\subsection{Cache Coverage Across Datasets}

Table~\ref{tab:dataset_coverage} aggregates the same full matrix by dataset. This view shows why the method is useful in some domains and fragile in others. FLEURS German, Spanish, French, Mandarin, and Japanese all show large average gains, indicating that cached LLM probabilities can be especially valuable for multilingual N-best lists with recoverable lexical or character-level alternatives. LibriSpeech test-other (LS) and noisy 5dB mixing LibriSpeech test-other (LS 5 dB mixing) show smaller but still positive average gains. AMI has a very low mean hit rate, which limits the use of exact source-cache lookup, but it can still benefit when the N-best list contains many recoverable errors. Thus, the dataset view supports the main thesis: usefulness depends jointly on N-best recoverability and cache coverage, not on teacher size alone.

\begin{table}[t]
\centering
\caption{Dataset-level cache behavior across teacher LLMs and available ASR models. WER/CER Values aggregate the $K=32$ rows in Table~\ref{tab:context_matrix}.}
\label{tab:dataset_coverage}
\begin{tabular}{lrlrr}
\toprule
Dataset & \#ASR*\#Teacher & Metric & Mean hit & Mean $\Delta_{32}$ \\
\midrule
AMI IHM & 12 & WER & 0.61 & -1.860 \\
FLEURS cmn & 6 & CER & 2.37 & -6.152 \\
FLEURS de & 6 & WER & 12.68 & -4.097 \\
FLEURS en & 12 & WER & 23.58 & -1.267 \\
FLEURS es & 6 & WER & 21.61 & -3.734 \\
FLEURS fr & 6 & WER & 21.94 & -5.112 \\
FLEURS ja & 6 & CER & 16.92 & -5.481 \\
LS & 12 & WER & 27.18 & -0.764 \\
LS 5dB mix & 12 & WER & 24.00 & -0.839 \\
\bottomrule
\end{tabular}
\end{table}

\section{Discussion}

The experiments show that cached LLM probability retrieval is useful because it turns an LLM into a reusable local prior for ASR hypotheses. The method is most effective when three conditions hold: the N-best list contains a better hypothesis, the first-pass score does not already select it, and the cache/backoff policy supplies enough LLM probability evidence to change the ranking. This explains the strongest results on Whisper-small and several FLEURS rows, as well as the consistent gains on wav2vec~2.0 and HuBERT CTC lists. In these settings, the recognizer produces recoverable alternatives, and the cached LLM score adds linguistic discrimination without retraining the recognizer.

The comparison with LoRA-GER clarifies the method's role. GER and KD train parameters that can generalize to unseen contexts, while cached retrieval does not train and is therefore sparser. However, this is also its advantage: it is simple to build, deterministic to inspect, and can be deployed as a local feature after constructing the offline cache. Table~\ref{tab:qwen_teacher_full} shows that this lightweight feature remains competitive with trained correction in many rows, even though GER has access to supervised adaptation. Thus, the proposed method should be viewed as a low-cost LLM rescoring layer and a complement to GER or distillation, not as a replacement for every trained correction system.

The context sweep provides a practical design rule. Direct LLM rescoring may benefit from long histories, but exact cached retrieval does not automatically. Under utterance-level N-best rescoring, $K>32$ mostly increases the number of distinct keys and cache-construction costs without improving ranking. This makes short-context caching attractive: it captures most of the useful local probability signals while keeping the cache small. When long-context information is needed, the selective policy is a better place to allocate online LLM computation, because it calls the teacher only for ambiguous misses that can affect the final decision.

Strong recognizers such as Whisper-large-v3 may leave little room for text-only rescoring, and some SpeechBrain CRDNN/RNNLM experiments show that acoustic plausibility can dominate linguistic preference. These observations do not negate the usefulness of cached retrieval; they define its operating regime. A stronger future system can combine this cache with semantic retrieval, learned backoff, better cache construction from domain logs, and explicit latency budgeting for selective online scoring.

\section{Conclusion}

In this paper, we presented cached LLM probability retrieval as a training-free LLM-based ASR rescoring method. A local teacher LLM is queried offline to build a probability cache, and an existing recognizer uses the cache online through lookup, shorter-context backoff, and optional selective direct scoring for important misses. Across full-data Whisper, self-supervised CTC, and SpeechBrain CRDNN/RNNLM N-best lists, the cached score improves first-pass ASR in 28 of 39 rows and is tied for the best non-oracle result in 25 rows. The gains are largest for Whisper-small, where all nine datasets improve, with an average best-cache reduction of 8.13\% absolute WER/CER, and they are also consistent for wav2vec~2.0 and HuBERT. The comparison with strong LoRA-GER baselines shows that trained correction remains valuable when supervised adaptation is available, but cached retrieval offers a simpler, inspectable alternative that can deliver substantial gains without fine-tuning. Finally, across the 78 context sweep experiments, exact cached retrieval does not require long contexts: no condition benefits from $K>32$, and $K=8$ is often sufficient. These findings support the use of cached LLM probability retrieval as a useful, lightweight component for ASR adaptation, especially when N-best hypotheses contain recoverable linguistic alternatives and local/offline deployment is preferred.

\section*{Acknowledgment and AI Disclosure}
This work was supported by JST BOOST, Japan Grant Number JPMJBY25F6. The authors used AI to (1) help generate source code for the experiments and (2) polish the English grammar and word usage of the manuscript. The idea is original, and the authors take full responsibility for the contents of the manuscript. 

\bibliographystyle{IEEEtran}
\bibliography{refarxiv}

\end{document}